\documentstyle[12pt]{article}
\begin{document}
\begin{titlepage}
\pagestyle{empty}
\baselineskip=21pt
\begin{center}
{\large{\bf Quantum Cosmology and the Constants of
Nature}\footnote{Talk given at the RESCEU International Symposium
``The Cosmological Constant and the Evolution of the Universe''}}
\end{center}
\vskip .1in
\begin{center}
Alexander Vilenkin

{\it Institute of Cosmology, Department of Physics and Astronomy}

{\it Tufts University, Medford, MA 02155, USA}

\vskip .1in

\end{center}
\vskip .5in
\centerline{ {\bf Abstract} }
\baselineskip=18pt
In models where the constants of
Nature can take more than one set of values, the cosmological wave
function $\psi$ describes an ensemble of universes with different
values of the constants.
The probability distribution for the constants can be determined with
the aid of the `principle of mediocrity' which asserts that we are a
`typical' civilization in this ensemble. I discuss the implications of
this approach for inflationary scenarios, the origin of density
fluctuations, and the cosmological constant.

\noindent

\end{titlepage}
\newpage
\baselineskip=18pt

\noindent{\bf 1.\quad Variable Constants}
\medskip
\nobreak

The observed values of the constants of Nature are conspicuously
non-random (`fine-tuned').  If particle masses are bounded by the Planck mass
$m_p$
and the coupling constants by 1, then a random selection would give
all masses $\sim m_p$ and all couplings $\sim 1$.  The vacuum energy
would then be $\rho_v \sim m_p^4$.  In contrast, some of the particle
masses are more than 20 orders of magnitude below $m_p$, and $\rho_v
\sim 10^{-120}m_p^4$.

There is a popular view that all constants will in the end be
determined from a unique logically consistent theory of everything.
However, the constants we observe depend not only on the fundamental
Lagrangian, but also on the vacuum state, which is likely not to be
unique.  For example, in higher-dimensional theories, the constants of
the four-dimensional world depend on the way in which the extra
dimensions are compactified.  The number of different
compactifications in superstring theories is believed to be
${\buildrel > \over \sim}
10^4$.  Moreover, Coleman (1988a) suggested
that all constants
appearing in sub-Planckian physics may become totally undetermined due to
Planck-scale wormholes connecting distant regions of spacetime.

It has been argued (Carter 1974; Carr \& Rees 1979; Barrow \& Tipler
1986)  that the values of the constants
are, to a large degree, determined by anthropic considerations: these
values should be consistent with the existence of conscious observers
who can wonder about them.  This `anthropic principle' has not been
particularly popular among cosmologists, since it appears to require
the existence of an ensemble of universes with different values of the
constants of Nature.  It should be pointed out, however, that one is
lead to the concept of such an ensemble if one adopts the view that
the universe should be described quantum-mechanically.
The approach I am going to describe here combines anthropic
considerations with quantum cosmology and inflationary scenario.
\bigskip

\noindent{\bf 2.\quad \bigskip Creation of Universes from Nothing}

\smallskip

\smallskip

The world view suggested by
quantum cosmology is that
small closed universes, with all possible values of the constants of
Nature, spontaneously nucleate out of nothing.  Here, `nothing' refers
to the absence of not only matter, but also of space and time
(Vilenkin 1982, 1986; Hartle \& Hawking 1983; Linde 1984).  After
nucleation, the universes enter a state of
inflationary (exponential) expansion.  It is driven by the potential
energy of a scalar field $\varphi$, while the field slowly `rolls
down' its potential $V(\varphi)$.  This vacuum energy eventually
thermalizes, and inflation is followed by the usual
radiation-dominated expansion.

The probability distribution for the initial states of nucleating
universes can be obtained from the cosmological wave function
$\psi(a,\varphi)$ which satisfies the Wheeler-DeWitt equation
\begin{equation}
\left[ {\partial^2 \over{\partial a^2}}-{1\over{a^2}}{\partial^2
\over{\partial\varphi^2}}-a^2 \left( 1-a^2
V(\varphi)\right)\right]\psi(a,\varphi)=0,
\label{WDW}
\end{equation}
Here, $a$ is the radius of the universe.  Eq.(\ref{WDW}) is
supplemented by the `tunneling' boundary condition which requires that
at $a\to\infty ~~~ \psi$ includes only outgoing waves (in other words,
that $\psi$ does not include any components describing universes
contracting from an infinitely large size).  Then one
finds that the probability for a universe to nucleate with a set of
constants $\{\alpha_j\}$ and with the initial value of the scalar
field between $\varphi$ and $\varphi+d\varphi$ is
\begin{equation}
dw_\alpha=\rho_\alpha(\varphi)d\varphi,
\end{equation}
where
\begin{equation}
\rho_\alpha(\varphi)\approx C_\alpha\exp\left(
-{3m_p^4\over{8V_\alpha(\varphi)}}\right)
\label{rho}
\end{equation}
and $C_\alpha\approx const$.  Here, the subscript `$\alpha$' is a
collective symbol for the constants $\{\alpha_j\}$.  The initial
radius of the nucleating universes is $a_0 =[V_\alpha(\varphi)]^{-1/2}$,
and the overall
normalization of (\ref{rho}) is determined by
\begin{equation}
\sum_\alpha \int\rho_\alpha(\varphi)d\varphi =1.
\end{equation}
For each set of $\{\alpha_j\}$, the distribution (\ref{rho}) is peaked
at the maximum of $V_\alpha(\varphi)$.  This is just the right initial
condition for inflation.

An alternative approach to boundary conditions for $\psi$, which is
based on Euclidean quantum gravity, has been developed by Hawking and
collaborators.  The expression for the probability
distribution obtained using this approach is the same as (\ref{rho})
but with a different sign in the exponential.  It is peaked at the
minimum of $V(\varphi)$ and does not favor initial states that lead to
inflation.  Here, I adopt the tunneling boundary condition.

We can think of the probability distribution $\rho_\alpha(\varphi)$ as
describing an ensemble of universes which can be called
`metauniverse'.  The probability that a
universe arbitrarily picked in this ensemble will have a particular
set of $\{\alpha_j\}$ is
\begin{equation}
w_\alpha =\int\rho_\alpha (\varphi)d\varphi\propto\exp\left(
-{3m_p^4\over{8V_\alpha^{(max)}}}\right),
\label{w}
\end{equation}
where $V_\alpha^{(max)}=max\{V_\alpha(\varphi)\}$.

Those who are not comfortable with the idea of other universes can
regard this ensemble of universes simply as a mathematical construct
for calculating probabilities, just as we do in ordinary
quantum mechanics.  In either case, the calculations are done {\it as
if} all the universes are `real', and the simplest view to adopt is
that they indeed are.
\bigskip

\noindent{\bf 3.\quad Principle of Mediocrity}
\medskip
\nobreak

It is quite possible that a universe randomly picked in the ensemble will be
unsuitable for life, and therefore the distribution (\ref{w}) is not
adequate for predicting the observed values of the constants.
Moreover, the number of civilizations in some of the universes may be
much greater than in the others, and this difference should also be
taken into account when evaluating the probabilities.\footnote{This
and the following sections are partly based on my papers (1995a,b,c).
Related ideas were discussed by Albrecht (1995) and by Garcia-Bellido
\& Linde (1995).}
The probability distribution of constants for a {\it civilization}
randomly picked in the metauniverse is
\begin{equation}
{\cal P}_\alpha =C^{-1} w_\alpha {\cal N}_\alpha,
\label{p}
\end{equation}
where ${\cal N}_\alpha$ is the average number of civilizations in a
universe with a set of constants $\{\alpha_j\}$ and $C=\sum_\alpha
w_\alpha{\cal N}_\alpha$ is a normalization constant.  ${\cal N}$ is
taken to be the total number of civilizations through the entire
history of the universe, rather than their number at some moment of
time.  (Comparing different universes at a given moment of time is not
a very meaningful procedure: some universes may recollapse while the
others are still expanding.  Besides, the results of such a
comparison are sensitive to the choice of the time
variable).

If we assume that our civilization is a `typical' inhabitant of the
metauniverse, then we `predict' that the constants of Nature in our
universe are somewhere near the maximum of the distribution (\ref{p}).
The assumption of being typical, which I called the `principle of
mediocrity', is a version of the
`anthropic principle'.  The motivation for a new name was
to emphasise the difference with the common practice of the anthropic
principle where one merely assigns vanishing probabilities to the values of
constants not suitable for life.

The number ${\cal N}$ can be expressed as
\begin{equation}
{\cal N}_\alpha ={\cal V}_\alpha \nu_{civ}(\alpha),
\label{n}
\end{equation}
where ${\cal V}_\alpha$ is the volume of the universe at the end of
inflation, and $\nu_{civ}(\alpha)$ is the average number
of civilizations originating per unit thermalized volume.
(More exactly, ${\cal V}$ is the volume of the 3-dimensional
hypersurface separating the inflating and thermalized regions of
spacetime).

The concept of `naturalness' that is commonly used to assess the plausibility
of elementary particle models is based on the assumption that the probability
distribution for the constants is nearly flat,
${\cal P}_\alpha \approx const$.
The principle of mediocrity gives a very different perspective on what is
natural and what is not.  It predicts that the constants $\{\alpha_j\}$ are
likely to be such that the product
\begin{equation}
{\cal P}_\alpha \propto w_\alpha{\cal V}_\alpha \nu_{civ}(\alpha)
\label{p1}
\end{equation}
is maximized.  The factors in this product have a strong (exponential)
dependence on $\{\alpha_j\}$, and the distribution ${\cal P}_\alpha$
can be
strongly peaked in some region of $\alpha$-space.

It should be emphasized that predictions of the principle of
mediocrity are not guaranteed to be correct.  After all, our
civilization may be special in some respects.  The predictions can be
expected to have only statistical accuracy.  That is, with a large
number of predictions, only few of them are likely to be wrong.
\medskip

\noindent{\bf 4.\quad Predictions for Finite Inflation}
\medskip
\nobreak

I will first assume that inflation has a finite duration.  The more
complicated case of eternal inflation will be discussed in Sec.5.

\medskip

\noindent{\it The inflaton potential}
\medskip
\nobreak

The volume factor ${\cal V}$ is given by ${\cal V}={\cal V}_0 Z^3$,
where ${\cal V}_0 \sim (GV_\alpha^{(max)})^{-3/2}$ is the initial
volume at
nucleation and $Z$ is the expansion factor during inflation.
The maximum of $Z$ is achieved by maximizing the highest value of the
potential $V_\alpha^{(max)}$, where inflation starts, and minimizing
the slope
of $V(\varphi)$: the field $\varphi$ takes longer to roll down for a
flatter potential.  [Note that the nucleation probability (5) is also
maximized for the highest allowed value of $V_\alpha^{max}$].

The cosmological literature abounds with remarks on the `unnaturally'
flat potentials required by inflationary scenarios.  With the
principle of mediocrity the situation is reversed: flat is natural.
Instead of asking why $V(\varphi)$ is so flat, one should now ask why
it is not flatter.
\medskip

\noindent{\it Low-energy physics}
\medskip
\nobreak

The `human factor' $\nu_{civ}(\alpha)$ may impose stringent
constraints on the constants $\{\alpha_j\}$.  We do not know what
other forms of intelligent life are possible, but the principle of
mediocrity favors the hypothesis that our form is the most common in
the metauniverse.  The conditions required for life of our type to exist
[the low-energy physics based on the symmetry group $SU(3)\times
SU(2)\times U(1)$, the existence of stars and planets, supernova
explosions] may then fix, by order of magnitude, the values of the
fine structure constant, and of electron, nucleon, and W-boson masses,
as discussed by Carter (1974), Carr \& Rees (1979) and Barrow \& Tipler
(1986).
\medskip

\noindent{\it Origin of structure}
\medskip
\nobreak

Superflat potentials required by the principle of mediocrity typically
give rise to density fluctuations which are many orders of magnitude
below the strength needed for structure formation.  This means that
the observed structures must have been seeded by some other mechanism.
An alternative mechanism is based on
topological defects: strings, global monopoles, and textures, which
could be formed at a symmetry breaking phase transition.  (For a
review of topological defects and their cosmological implications see,
e.g., Vilenkin \& Shellard 1994).  The required symmetry breaking scale
for the defects is $\eta\sim
10^{16}~GeV$.  With `natural' (in the traditional sense) values of the
couplings, the transition temperature $T_c\sim\eta$ is much
higher than the  thermalization temperature, and no
defects are formed after thermalization.  It is possible for the phase
transition to occur during inflation, but the resulting defects are
inflated away, unless the transition is sufficiently close to the end
of inflation.  To arrange this requires some fine-tuning of the
constants.  However,
since the dependence of the volume factor ${\cal V}$ on the duration
of inflation is exponential, we expect that the gain in the volume
will more than compensate for the decrease in `$\alpha$-space' due to
the fine-tuning.\footnote{We note also that in some supersymmetric models the
critical temperature of superheavy string formation can `naturally' be
as low as $m_W$ (Lazarides {\it et. al.} 1986).}
Moreover, it has been recently shown (Kofman {\it et. al.} 1995) that
large fluctuations of the inflaton field $\varphi$ prior to
thermalization can result in the formation of superheavy defects, even
in models with a low thermalization temperature.

Another possibility is to use more complicated models of inflation,
such as `hybrid' inflation (Linde 1994), which involve several scalar
fields and can give reasonably large density fluctuations even when
the potentials are very flat in some directions in the field space.

The symmetry breaking scale $\eta\sim 10^{16}~GeV$ for the defects is
suggested by observations, but we have not explained why this
particular scale has been selected.  The value of $\eta$ determines
the amplitude of density fluctuations, which in turn determines the
time when galaxies form and the matter density in the galaxies.
Since these parameters certainly affect
the chances for civilizations to develop, it is quite possible that
$\eta$ is significantly constrained by the anthropic factor
$\nu_{civ}(\alpha)$.  It would therefore be interesting to study how
structure formation would proceed in a universe with a very different
amplitude of density fluctuations (and/or a very different baryon
density).  Some  steps in this
direction have been made by Rees (1980, 1995).
\medskip

\noindent{\it The cosmological constant}
\medskip
\nobreak

An anthropic
bound on the cosmological constant has been first discussed by Weinberg
(1987).  In a spatially flat universe with a positive vacuum
energy density $\rho_v$, gravitational clustering effectively stops at
a redshift
\begin{equation}
1+z_v \sim (\rho_v /\rho_{m0})^{1/3},
\end{equation}
when $\rho_v$ becomes comparable to the matter density $\rho_m$.  At
later times, the vacuum energy dominates, and the universe enters a
deSitter stage of exponential expansion.  A bound on $\rho_v$ can be
obtained by requiring that it does not dominate before at least a single
galaxy had a chance to form.  There is evidence for the
existence of quasars and protogalaxies as early as $z\sim 4$, and
Weinberg argued that the
anthropic principle, interpreted in this way,
cannot rule out vacuum energy domination at $z
{\buildrel< \over \sim} 4$.  This corresponds to the bound
\begin{equation}
\rho_v /\rho_{m0}{\buildrel < \over \sim} 100,
\label{wbound}
\end{equation}
which falls short of
the observational upper bound (see, e.g., Turner 1995),
\begin{equation}
(\rho_v/\rho_{m0})_{obs} {\buildrel < \over \sim} 5,
\label{obound}
\end{equation}
by a factor $\sim 10$.

On the other hand, the principle of mediocrity suggests that we look
not for the value of $\rho_v$ that makes galaxy formation barely
possible, but for the value maximizing the number of inhabitable
stellar systems.  As a rough measure of the latter, we
can use the fraction of baryonic matter, $f(\rho_v)$, that ends up in
giant galaxies which contain most of the luminous stars in the
Universe.
Dwarf galaxies with masses $M \ll 10^{11}M_\odot$ are vulnerable to losing
much of their gas through winds driven by supernovae from
the first generation of star formation (Dekel \& Silk 1986; Babul \&
Rees 1992; Nath \& Chiba 1995).  (Note that
planetary systems and life do not form before the heavy elements
produced in the first-generation stars are dispersed in supernova
explosions).  In bottom-up
structure formation scenarios, where giant galaxies are formed
mainly by aggregation of smaller components, the chances for life to
evolve may be greatly diminished if the vacuum energy dominates before
the formation of giant galaxies.\footnote{Observationally, there has
been no strong evolution of giant galaxies
since $z\sim 1$ (see, e.g., Ellis 1995).  At the same time, there has been a
marked decrease in the numbers of dwarf galaxies since $z\sim 0.5-1$.
At least part of this decrease could be due to the absorbtion of
dwarfs by larger
galaxies, and thus it is conceavable that $f(\rho_v)$ grew, say, by a
factor of a few at $z{\buildrel < \over \sim}1$.  The width of the distribution
(\ref{lambda}) would then be comparable to the observational bound
(\ref{obound}).}  In top-down scenarios, such as hot dark matter with
cosmic string seeds, first structures can form at high redshifts, but
the bulk of galaxy formation occurs at $z\sim 1-2$, and thus the most
probable values of $\rho_v$ are in the range $z_v {\buildrel < \over
\sim} 1$ and $\rho_v /\rho_{m0} {\buildrel < \over \sim} 10$.

In both cases, the function $f(\rho_v)$ decreases from
$f(0) \sim 1$ to negligibly small values at $\rho_v
/\rho_{m0} \gg 100$.
At the same time, one expects the nucleation probability and the
volume factor ${\cal V}(\rho_v)$ to vary on a much greater scale (say,
$\sim V_\alpha^{max}$), and therefore to be essentially constant in
the range where $f(\rho_v)$ is substantially different from
zero.\footnote{A nucleation probability with an extremely sharp peak
at $\rho_v = 0$ was obtained by Baum (1984), Hawking
(1984), and Coleman (1988b).  However, all these
papers are based on Euclidean quantum gravity which has serious
problems.  For a discussion of the problems, see Fischler {\it
et. al.} (1989).}  Hence,
the probability distribution for $\rho_v$, with all other constants
fixed, is
\begin{equation}
d{\cal P} \propto f(\rho_v)d\rho_v ,
\label{lambda}
\end{equation}
where I have assumed that $\rho_v$ has a continuous
spectrum and that it can vary independently of other constants.

A somewhat similar interpretation of the anthropic principle has been
independently used by
Efstathiou (1995) who defined the probability density for
$\rho_v$ by estimating the number of giant ($L^*$) galaxies per
photon, $N_g$, in a cold
dark matter model with different values of $\rho_v$.  An
important difference between his approach and mine is that Efstathiou
calculated $N_g (\rho_v)$ at the moment of time corresponding to a
fixed (present) value of the Hubble parameter $H_0$, while my
suggestion (Vilenkin 1995a,b) is to use the number of civilizations
throughout the history of the universe, which would correspond to
calculating the asymptotic value of $N_g (\rho_v)$ as $t\to\infty$.
As noted by Rees (private communication), by fixing $H_0$ one eliminates all
values of $\rho_v$ smaller than $3H_0^2/8\pi G$, and thus the analysis
of $N_g$ at a fixed $H_0$ is not suitable for explaining the smallness
of $\rho_v$.

For negative values of $\rho_v$, the scale factor behaves as
$a(t)\propto [\sin (\pi t/t_c)]^{2/3}$,
where $t_c=(\pi/6|\rho_v|)^{1/2}$, so that the universe recollapses on
a timescale $t_c$.  The width of the probability distribution for
$\rho_v <0$ is determined by requiring that the
universe does not recollapse while stars are still shining and new
civilizations are being formed.  This gives a width comparable to
(\ref{obound}).
\medskip

\noindent{\bf 5.\quad Predictions for Eternal Inflation}
\medskip
\nobreak

I have assumed so far that inflation has a finite duration, so that
the thermalized volume ${\cal V}$ and the number of civilizations
${\cal N}$ are both finite.  This, however, is not generally the case.
The evolution of the inflaton field $\varphi$ is influenced by quantum
fluctuations, and as a result thermalization does not occur
simultaneously in different parts of the universe.  In many models it
can be shown that at any time there are parts of the universe that are
still inflating (Vilenkin 1983; Linde 1986).  Conclusions of
Section 4 are directly applicable only if inflation is finite for all
the allowed values of the constants.

In an eternally inflating universe, the thermalization volume ${\cal
V}$ is infinite and has to be regulated.  If one simply cuts it off by
including only parts of the volume that thermalized prior to some
moment of time $t_c$, with the same value of $t_c$ for all universes,
then one finds that the results are extremely sensitive to the choice
of the time coordinate $t$.  For example, cutoffs at a fixed proper
time and at a fixed scale factor $a$ give drastically different
results (Linde {\it et. al.} 1994).  An alternative
procedure (Vilenkin 1995; Winitzki \& Vilenkin 1995), which is
free of this problem, is to
introduce a cutoff at the time when all but a small fraction
$\epsilon$ of the initial (co-moving) volume of the universe has
thermalized.  The value of $\epsilon$ is taken to be the same for all
universes, but the corresponding cutoff times $t_c$ are generally
different.  The limit $\epsilon \to 0$ is taken after calculating the
probability distribution ${\cal P}_\alpha$.  It can be shown
that the resulting distribution is not sensitive to
the choice of $t$.

I will omit the calculation of the regularized volume ${\cal V}$
and even the rather lengthy expression for ${\cal V}$
obtained as a result of that calculation.  The essence of the result
can be expressed as
\begin{equation}
{\cal V}\propto\epsilon^{-d/(d-3)}Z^3.
\label{vol}
\end{equation}
Here, $Z$ is the expansion factor during the slow-roll phase of
inflation, when quantum fluctuations are small, and $d<3$ has the
meaning of the fractal dimension of the inflating region.

In the limit $\epsilon\to 0$, non-vanishing probabilities are
obtained only for $\{\alpha_j\}$ corresponding to the largest value
of $d$,
\begin{equation}
d(\alpha)=max.
\label{gamma}
\end{equation}
The fractal dimension $d$ increases as the potential $V(\varphi)$
becomes flatter, and thus the condition (\ref{gamma})
tends to select maximally flat potentials.

It is possible that the condition (\ref{gamma}) selects a unique set
of $\{\alpha_j\}$.  Then all constants of Nature can, at least in
principle, be predicted with 100\% certainty.  On the other hand, if
the maximum of $d$ is strongly degenerate, then
Eq.(\ref{gamma}) selects a large subset of all $\{\alpha_j\}$.
All values of $\alpha$ not in this subset have a vanishing
probability, and the probability distribution within the subset is
proportional to $w_\alpha Z^3_\alpha\nu_{civ}(\alpha)$ [see
Eq.(\ref{p1})].  The probability maximum is then determined by the same
considerations as in the case of finite inflation.

This situation will arise, for example, if the potential
$V(\varphi)$ has a large number of minima, parametrized by some subset
of the constants $\{\beta_j\} \subset \{\alpha_j\}$.
Thermalization will then occur to different minima in different
parts of the universe.  The regularized volume in this case is still
given by Eq.(\ref{vol}), but now $d$ has the same value everywhere
and is independent of $\{\beta_j\}$ (Linde {\it et. al.} 1994).
{}From Eq.(\ref{vol}), ${\cal V}_\beta \propto Z_\beta^3$, and the
corresponding probabilities are
\begin{equation}
{\cal P}_\beta \propto Z_\beta^3\nu_{civ}(\beta).
\label{pcms}
\end{equation}
The probability distribution (\ref{pcms}) has the same dependence on
the slow-roll expansion factor $Z$ and on the anthropic factor
$\nu_{civ}$ as we found in the case of finite inflation.
The predictions for $\{\beta_j\}$ are, therefore, essentially the same
(see Section 4).

Another possibility is that the `constants' of low-energy physics are
affected by some very weakly coupled fields.  These could be
the moduli fields of superstring theories, one example being the
dilaton which determines the value of the Newton's constant $G$.  Such
fields should be included in $\{\beta_j\}$ as continuous
variables.\footnote{The probability
distribution for a Brans-Dicke field (which is similar
to the dilaton) was discussed,
using a different approach, by Garcia-Bellido {\it et. al.} (1994, 1995).}
\medskip

\noindent{\bf 6.\quad Conclusions}
\medskip
\nobreak

When quantum mechanics is applied to the entire universe, we are
inevitably lead to the concept of an ensemble of universes, with a
probability distribution for different initial conditions and for
different constants of Nature.  The principle of mediocrity asserts
that we are typical inhabitants of this ensemble.  The values of the
constants suggested by this principle are the ones that give a very
flat inflaton potential, a non-negligible cosmological constant, and
density fluctuations seeded either by topological defects or by
quantum fluctuations in models like `hybrid' inflation.  We can expect
to find these features in our own universe, provided that they are
consistent with the spectrum of the allowed constants $\{\alpha_j \}$.
The spectrum of $\{\alpha_j\}$ will hopefully be determined from the
fundamental particle theory, and until then no reliable predictions
can be made for the values of the constants.  However, this
preliminary analysis suggests that the predicted values may be very
different from the choices considered `natural' by particle
physicists.
\medskip

\noindent{\bf Acknowlegements}
\medskip
\nobreak

It is a pleasure to thank Katsuhiko Sato and all organizers of the
Symposium for their warm hospitality at the University of Tokyo.
I benefited from discussions with G.Efstathiou, M.Fukugita, A.Linde,
M.Rees, D.Spergel, and S.White, although most of them will probably
not subscribe to the views presented in this paper.  This research was
supported in part by the National Science Foundation (USA).
\medskip

\noindent{\bf References}
\medskip
\nobreak

1.  Albrecht, A., 1995, in {\it The Birth of the Universe and
Fundamental Forces}, ed. by F. Occhionero, Springer-Verlag.

2.  Babul, A. and Rees, M.J., 1992, {\it M.N.R.A.S.} {\bf 255}, 346.

3.  Barrow, J.D. and
Tipler, F.J., 1986, {\it The Anthropic Cosmological Principle},
Clarendon, Oxford.

4.  Baum, E., 1984, {\it Phys. Lett.} {\bf B133}, 185.

5.  Carr, B.J. and Rees,
M.J., 1979, {\it Nature (London)} {\bf 278}, 605.

6.  Carter, B., 1974, in {\it I.A.U. Symposium, Vol.} {\bf 63}, ed. by
M.S. Longair, Reidel, Dordrecht; 1983, {\it
Philos. Trans. R. Soc. London} {\bf A310}, 347.

7.  Coleman, S., 1988a, {\it Nucl. Phys.} {\bf B307}, 867.

8.  Coleman, S., 1988b, {\it Nucl. Phys.} {\bf B130}, 643.

9.  Dekel, A. and Silk, J., 1986, {\it Ap. J.} {\bf 303}, 39.

10. Efstathiou, G., 1995, {\it Mon. Not. R. Astron. Soc.} {\bf 274},
L73.

11. Ellis, R.S., 1995, The morphological evolution of galaxies,
astro-ph/9508044.

12. Fischler, W., Klebanov, I., Polchinski, J. and Susskind, L., 1989,
{\it Nucl. Phys.} {\bf B237}, 157.

13. Garcia-Bellido, J. Linde, A.D. and Linde, D.A., 1994, {\it Phys. Rev.}
{\bf D50}, 730.

14. Garcia-Bellido, J. and Linde, A.D., 1995, {\it Phys. Rev.} {\bf D51},
429.

15. Hartle, J.B. and Hawking, S.W., 1983, {\it Phys. Rev.} {\bf D28},
2960

16. Hawking, S.W., 1984, {\it Phys. Lett.} {\bf B134}, 403.

17. Kofman, L., Linde, A.D. and Starobinsky, A.A., 1995, Non-thermal phase
transitions after inflation, hep-th/9510119.

18. Lazarides, G., Panagiotakopoulos, C. and Shafi, Q., 1986, {\it
Phys. Rev. Lett.} {\bf 56}, 432.

18. Linde, A.D., 1984, {\it
Lett. Nuovo Cim.} {\bf 39}, 401

19. Linde, A.D., 1986, {\it Phys. Lett.} {\bf B175}, 395.

20. Linde, A.D., 1994, Phys. Rev. {\bf D49}, 748.

21. Linde, A.D., Linde, D.A. and Mezhlumian, A., 1994, {\it
Phys. Rev.} {\bf D49}, 1783.

22. Nath, B.B. and Chiba, M.,1995, astro-ph/9505081.

23. Rees, M.J., 1980, {\it Phys. Scripta} {\bf 21}, 614.

24. Rees, M.J., 1995, these Proceedings.

25. Turner, E., 1995, these Proceedings.

26. Vilenkin, A., 1982, {\it Phys. Lett.} {\bf 117B}.

27. Vilenkin, A., 1983, {\it Phys. Rev.} {\bf D27}, 2848.

28. Vilenkin, A., 1986, {\it Phys. Rev.} {\bf D33}, 3560.

29. Vilenkin, A., 1995a, {\it Phys. Rev. Lett.} {\bf 74}, 846.

30. Vilenkin, A., 1995b, Predictions
from quantum cosmology (Erice lectures), gr-qc/9507018.

31. Vilenkin, A., 1995c, {\it Phys. Rev.} {\bf D52}, 3365.

32. Vilenkin, A. and Shellard, E.P.S., 1994,
{\it Cosmic Strings and Other
Topological Defects}, Cambridge University Press, Cambridge.

33. Weinberg, S., 1987, {\it Phys. Rev. Lett.} {\bf 59}, 2607.

34. Winitzki, S. and Vilenkin, A., 1995, Uncertainties of predictions in
models of eternal inflation, gr-qc/9510054.

\end{document}